\documentclass[conference]{IEEEtran}
\IEEEoverridecommandlockouts
\usepackage{cite}
\usepackage{amsmath,amssymb,amsfonts}
\usepackage{algorithmic}
\usepackage{graphicx}
\usepackage{textcomp}
\usepackage{xcolor}
\usepackage{hyperref}
\def\BibTeX{{\rm B\kern-.05em{\sc i\kern-.025em b}\kern-.08em
    T\kern-.1667em\lower.7ex\hbox{E}\kern-.125emX}}
\begin{document}


\title{Predicting Workload in Virtual Flight Simulations using EEG Spectral and Connectivity Features (Including Post-hoc Analysis in Appendix)} 

\author{\hspace{-2em}\IEEEauthorblockN{1\textsuperscript{st} Bas Verkennis}
\IEEEauthorblockA{\textit{Department of Cognitive Science and Artificial Intelligence\hspace{1em}} \\
\hspace{-2em}\textit{Tilburg University}\\
\hspace{-2em}Tilburg, Netherlands \\
\hspace{-2em}b.verkennis@tilburguniversity.edu}
\and
\IEEEauthorblockN{2\textsuperscript{nd} Evy van Weelden*}
\IEEEauthorblockA{\textit{Department of Cognitive Science and Artificial Intelligence} \\
\textit{Tilburg University}\\
Tilburg, Netherlands \\
e.vanweelden@tilburguniversity.edu*}
\IEEEauthorblockA{*Corresponding author.}
\and
\IEEEauthorblockN{3\textsuperscript{rd} Francesca L. Marogna}
\IEEEauthorblockA{\textit{Department of Cognitive Science and Artificial Intelligence} \\
\textit{Tilburg University}\\
Tilburg, Netherlands \\
f.l.marogna@tilburguniversity.edu}
\and
\hspace{-4em}
\IEEEauthorblockN{4\textsuperscript{th} Maryam Alimardani}
\IEEEauthorblockA{\textit{Faculty of Science, Computer Science\hspace{4em}} \\
\hspace{-5em}\textit{Vrije Universiteit Amsterdam}\\
\hspace{-5em}Amsterdam, Netherlands \\
\hspace{-5em}m.alimardani@vu.nl}
\and
\IEEEauthorblockN{5\textsuperscript{th} Travis J. Wiltshire}
\IEEEauthorblockA{\textit{Department of Cognitive Science and Artificial Intelligence} \\
\textit{Tilburg University}\\
Tilburg, Netherlands \\
t.j.wiltshire@tilburguniversity.edu}
\and
\hspace{1em}
\IEEEauthorblockN{6\textsuperscript{th} Max M. Louwerse}
\IEEEauthorblockA{\hspace{2em}\textit{Department of Cognitive Science and Artificial Intelligence} \\
\hspace{2em}\textit{Tilburg University}\\
\hspace{2em}Tilburg, Netherlands \\
\hspace{2em}m.m.louwerse@tilburguniversity.edu}
}

\maketitle
\IEEEpubidadjcol

\begin{abstract}
Effective cognitive workload management has a major impact on the safety and performance of pilots. Integrating brain-computer interfaces (BCIs) presents an opportunity for real-time workload assessment. Leveraging cognitive workload data from high-fidelity virtual reality (VR) flight simulations allows for dynamic adjustments to training scenarios. While prior studies have predominantly concentrated on EEG spectral power for workload prediction, delving into intra-brain connectivity may yield deeper insights. This study assessed the predictive value of EEG spectral and connectivity features in distinguishing high vs. low workload periods during simulated flight in VR and Desktop conditions. Using an ensemble approach, a stacked classifier was trained to predict workload from the EEG signals of 52 participants. Results showed that the mean accuracy of the model incorporating both spectral and connectivity features improved by 28\% compared to the model that solely relied on spectral features. Further research on other connectivity metrics and deep learning models in a large sample of pilots is essential to validate the potential of a real-time workload-prediction BCI. This could contribute to the development of an adaptive training system for safety-critical operational environments.
\end{abstract}


\section{Introduction}
Ensuring pilot safety and performance during flight training depends on balancing cognitive effort and task demands \cite{vanweelden2022comparing}. Cognitive workload levels that are too high are detrimental to individual attention, decision-making, and other cognitive processes in ways that can lead to errors and potential flight hazards \cite{vanweelden2022aviation, dehais2019, arico2017}. On the other hand, cognitive workload levels that are too low may lead to attention deficits and performance decrements \cite{young2014}. Measuring cognitive workload in real time would allow for adaptive virtual reality (VR) systems that minimize pilot fatigue and optimize resource allocation during training \cite{arico2016, prinzel2000}. Real-time assessment of a pilot’s cognitive workload can be monitored by brain-computer interfaces (BCIs). These systems use a pilot's brain signals, for instance, recorded by electroencephalography (EEG) \cite{arico2016, neuroergonomics2019, zander2011, li2011}. Despite the recent advancements of BCIs for application in the aviation field, there are still challenges to overcome before these systems can be reliably used in real-world applications. These challenges include individual differences in brain activity, external factors such as background noise and muscle artifacts, and the identification of appropriate EEG features indicative of workload. The identification of EEG

\newcommand\blfootnote{ 
\begingroup \renewcommand
\thefootnote{} 
\endgroup } 

\noindent\rule{\linewidth}{0.4pt} 
\noindent \blfootnote{\textit{This research is funded by the MasterMinds project, as part of the RegionDeal Mid- and West-Brabant, and co-funded by the Ministry of Economic Affairs and the Municipality of Tilburg.} \\\\ \textbf{The official version of this paper is published in IEEE with DOI: \href{https://doi.org/10.1109/AIxVR63409.2025.00019}{10.1109/AIxVR63409.2025.00019}.} }



\noindent features associated with different mental states remains a key focus for research and development in the field. 

Previous studies have primarily relied on EEG spectral features to assess and predict workload \cite{vanweelden2022aviation, diazpiedra2019, dehais2017, feltman2021, jeong2019}. 
Spectral power refers to the distribution of energy or strength of neural activity within different frequency bands in the EEG signal \cite{7307237}. This information is crucial for understanding how different brain regions are activated during various tasks or cognitive processes. Notably, alpha power has been associated with increased arousal, resource allocation, or workload \cite{Ray1985, Fink2005, Pfurtscheller1996, brouwer2012}. Other frequency bands, including beta, delta, and gamma, have also been shown to be responsive to workload in air traffic control and simulated multi-task operations \cite{Brookings1996, Laine2002}. Moreover, research has consistently demonstrated that high workload conditions in flight simulation correspond to an increase in beta and theta power in the frontal region, as well as an increase in alpha and theta power in the parieto-occipital regions of the brain. Conversely, low workload conditions are associated with decreased beta and theta power in the frontal area and a further decrease in theta power in the central and parietal cortices. These findings \cite{vanweelden2022aviation, prinzel2000, diazpiedra2019, dehais2017, dussault2005, feltman2021, jeong2019, wright2005}, albeit varying by task, collectively highlight the significance of EEG spectral features in assessing cognitive workload levels in the aviation field. 

While spectral-based features are promising for workload classification, emerging evidence suggests that connectivity features may hold additional insights related to workload. \textit{Functional connectivity} quantifies how different regions of the brain interact to process information, potentially providing a more holistic view of cognitive processes \cite{leeuwis2021functional}. Unlike spectral power, functional connectivity captures the synchronization and coordination of brain activity, which is particularly valuable when capturing the broader network dynamics involved in cognitive processing. Kakkos et al. \cite{kakkos2019}, for instance, investigated alterations in functional brain networks during simulated flight experiments with varying predefined mental workload levels. Their study showed that it is possible to classify three levels of mental workload with a small subset of Phase Lag Index (PLI) features. Guan et al. \cite{guan2022} obtained functional connectivity in the fronto-parietal region using Phase Locking Value (PLV) and showed this connectivity to reflect changes in mental workload during various tasks. Thus, connectivity-based metrics may also prove useful for workload classification. 

Despite the current findings, the combination of both EEG spectral and connectivity features has rarely been investigated. One of the reasons is that EEG datasets are normally small-scale and extracting various features from the data creates the curse of dimensionality \cite{lotte2018}. To mitigate the risk of overfitting, many researchers turn to specific feature selection methods. These techniques serve a dual purpose: not only do they reduce the overall number of features aiding interpretability, but they also play a pivotal role in accurately pinpointing the most significant features for the classification task. For instance, Kakkos et al. \cite{kakkos2019} utilized Recursive Feature Elimination (RFE) with correlation bias reduction, while Taheri Gorji et al. \cite{taheri2023} conducted a comparison of RFE and LassoCV. These practices indicate that this selection process is instrumental in developing an accurate model for workload assessment. 

Additionally, supervised classification of workload requires accurate labeling of the EEG data. Prior studies have primarily assessed workload in predetermined workload scenarios, assuming that one task induces a higher workload than another task for all participants \cite{kakkos2019, taheri2023, gateau2018}. However, these methods have proven difficult to generalize across different studies, subjects, and domains \cite{grier2015}. In contrast, only a small number of studies predicted \cite{Nittala2018} or estimated \cite{yim2022} workload using self-reported levels of workload. The NASA-TLX \cite{hart1988} is a well-established instrument for assessing subjective workload across multiple constructs, including Mental Demand, Physical Demand, Temporal Demand, Performance, Effort, and Frustration. Traditionally used as a within-subject comparison tool \cite{grier2015}, NASA-TLX allows participants to provide their perceived level of workload using a Likert scale. Therefore, this instrument can be employed to collect subject-specific labels of workload for the EEG data to enhance the robustness of classification. 

By integrating BCIs with effective workload assessment techniques within VR flight simulations, we can create adaptive training systems that respond in real time to a pilot’s cognitive workload. This integration not only promises to improve the accuracy of workload assessments but also presents exciting opportunities for the future of flight training, enhancing pilot performance and safety through tailored experiences. The current study aims to contribute to these ongoing efforts by examining whether EEG fronto-parietal connectivity can serve as a reliable predictor of workload during simulated flight. For this purpose, we compared a Baseline model trained with only EEG spectral features to a Connectivity model that was trained with a selection of both spectral and connectivity features to discriminate between low and high workload conditions as reported by NASA-TLX. To obtain connectivity features, we computed the PLV, which measures phase synchrony between two EEG signals. To the best of our knowledge, PLV has never been examined before for workload estimation in the aviation domain. 

\begin{figure*}[htbp]
\centerline{\includegraphics[width=0.8\textwidth]{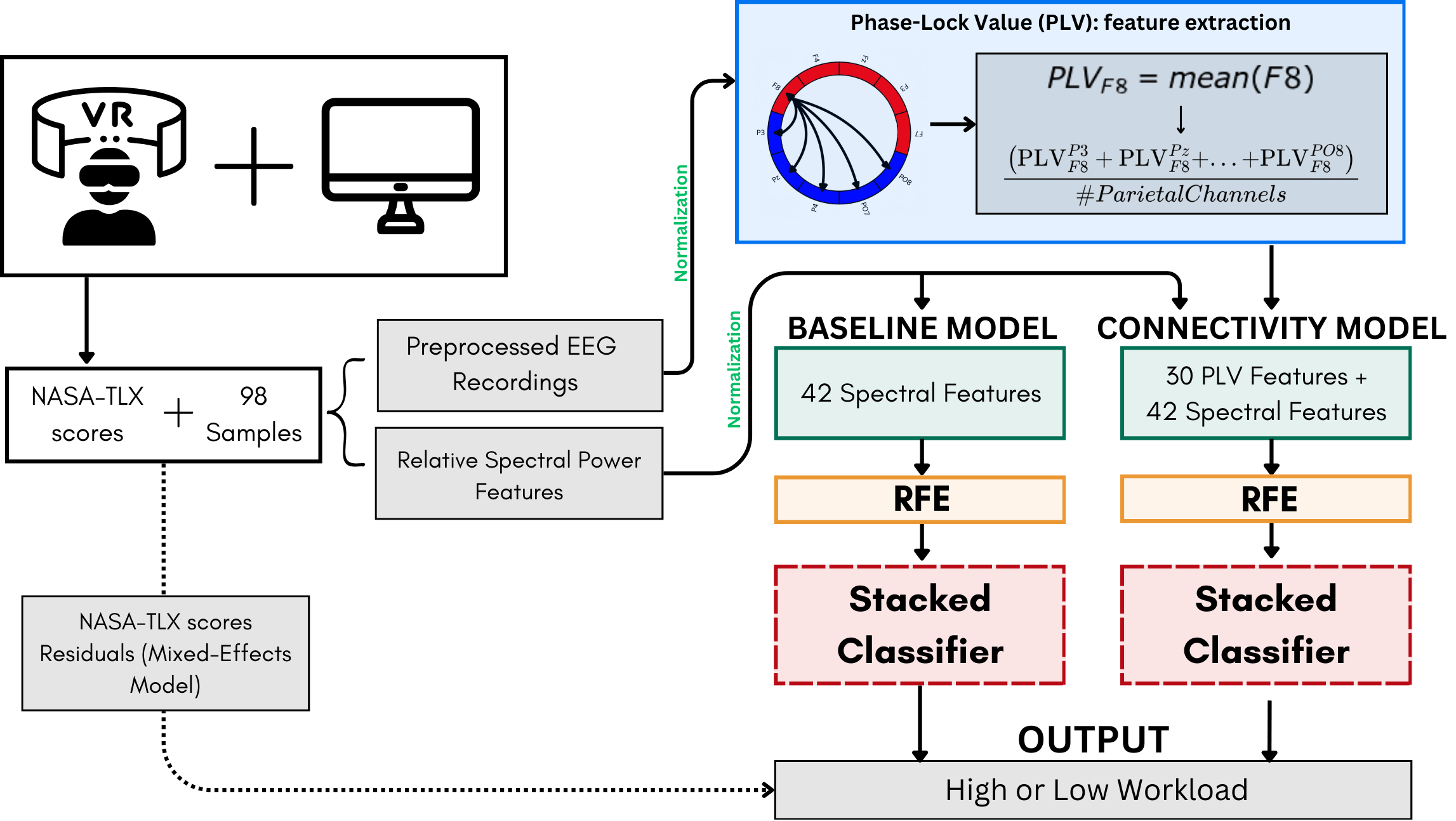}}
\caption{Overview of the BCI classification pipeline. EEG = electroencephalogram; PLV = Phase Locking Value; RFE = Recursive Feature Elimination; Stacked Classifier = ensemble model combining Random Forest, Logistic Regression, and SVM as base models, with SVM used as the meta-model; NASA-TLX = NASA Task Load Index.}
\label{fig:bciloop}
\end{figure*}

\section{Methods}
The processing steps and classification pipeline are visualized in Figure \ref{fig:bciloop}. The complete pipeline was constructed in Python, using the packages \textit{MNE-Python} \cite{gramfort2013} and \textit{sklearn} \cite{pedregosa2011}. Code is available at \cite{verkennis2024github}.

\vspace*{-3pt}
\subsection{Dataset Description}
The dataset used in this study was collected and pre-processed by van Weelden et al. \cite{vanweelden2024b}. The dataset includes EEG signals from 53 novice (non-pilot) subjects engaged in two flight simulations. Four participants were deemed outliers and were subsequently excluded from the analysis. 

The authors of \cite{vanweelden2024b} initially investigated whether an EEG-based index of workload and task engagement could predict performance during varying flight simulations. Each participant in the study engaged in flight tasks across two flight simulations. In one simulation, they utilized VR technology, which immersed participants in a high-fidelity 3D environment. In the other simulation, they used a standard 2D interface displayed on a regular computer screen. This allowed for a comparison between the two simulations (3D vs. 2D) in terms of user experience and performance. Following each condition, participants completed questionnaires, including NASA-TLX, which quantifies subjective workload on a scale from 0 to 100 \cite{hart1988}. 

The original dataset consisted of 32 electrodes arranged according to the 10-20 system, from which 14 electrodes (F7, F3, Fz, F4, F8, T7, Cz, T8, P3, Pz, P4, PO7, PO8, and Oz) mostly situated in the frontal, temporal, and parietal areas were chosen for the analysis. This selection was based on our earlier literature review \cite{vanweelden2022aviation} of brain areas that have been associated with workload estimation. From this selection, we extracted relative spectral power features and calculated fronto-parietal Phase Locking Values (PLV) in the theta, alpha, and beta frequency bands (see details in D. Feature Extraction). The EEG data encompassed two distinct event phases: the `baseline' and the `test' phases. The baseline EEG recordings were used for normalizing the signals in the test phase to minimize individual differences. 

\subsection{Binary Classification of Workload using NASA-TLX}
To conduct binary classification distinguishing between low and high workload states, we employed a median split of the NASA-TLX scores after we combined data from both Desktop and VR conditions to increase the sample size. Combining these two datasets could introduce biases, such as carryover effects or ``identity confounding", where consecutive tasks or individual differences might influence performance \cite{sutter2011, chaibubneto2019}. A post-hoc analysis (reported in Appendix~\ref{AppendixA}) found no evidence of such biases, supporting the validity of merging the datasets. Initially, we planned to use a threshold of 50 to split NASA-TLX scores, marking the midpoint of the NASA-TLX scale. However, this led to an imbalance between classes, with 65 samples reporting NASA-TLX scores at or below 50 and 33 samples reporting scores above 50. This imbalance could skew the model’s performance and generalizability, particularly affecting the minority class \cite{roy2019}. Moreover, this threshold-based approach, while straightforward, fails to fully consider NASA-TLX’s primary design as a within-subject tool, which captures subjective perceptions of workload rather than providing a universal metric for between-subject comparisons \cite{grier2015}. Therefore, applying a fixed threshold could introduce potential issues, given the inherent variability in workload ratings between different subjects. In response to this, residuals from a regression model were utilized to account for subject-specific variability \cite{chaibubneto2019, grier2015}, effectively addressing individual baseline differences and mitigating their potential impact on the analysis (reported in Appendix~\ref{AppendixB}, which discusses the mixed-effects regression modeling approach). 

Moreover, we utilized a subset of NASA-TLX subscales rather than relying on the overall NASA-TLX score \cite{hart2006, hertzum2021, tubbs-cooley2018, braarud2020}. This approach enabled us to pinpoint sources of workload and performance issues with greater precision (reported in Appendix~\ref{AppendixC}, which reviews the correlations between subscales and their implications for workload assessment). 
To address the issues of class imbalance and sample size, a pragmatic solution was implemented. Instead of resorting to data augmentation techniques, which might amplify overfitting concerns \cite{roy2019}, this study opted for combining data from both Desktop and VR conditions and employing a median split of the NASA-TLX scores for workload classification. Reusing participants' EEG data for data augmentation is a common strategy in BCI and machine learning (ML) studies to enhance the generalization of analysis \cite{roy2019}. This resulted in 98 independent samples that were divided into two equal classes after the median split. 

\subsection{Data Pre-processing}
The Signal Processing Toolbox \cite{mathworks} and the EEGLAB open-source toolbox \cite{delorme2004} in MATLAB R2021a were used to pre-process the raw EEG data, addressing signal artifacts \cite{delorme2004}. This pre-processing approach, as outlined in \cite{vanweelden2024b}, involved applying a low-pass filter with a 45Hz cutoff to remove high-frequency noise, particularly those associated with muscle movements \cite{ferrante2022, muthukumaraswamy2013}. In addition, Independent Component Analysis (ICA) was performed to identify and manually reject independent components related to eye movements and other artifacts. 

\renewcommand{\arraystretch}{1.5} 
\begin{table*}[htbp]
\fontsize{8}{10}\selectfont
\caption{Predictive Models Performance Metrics}
\resizebox{\textwidth}{!}{%
\begin{tabular}{|c|c|c|c|}
\hline  \hline  
\textbf{Model} 
& \textbf{Ranked Features} 
& \multicolumn{2}{c|}
{\textbf{Performance Metrics}} 
\\ \cline{3-4}
 & & 
 \textbf{Accuracy} (Mean $\pm$ SD, \%) & \textbf{F1-Score} (Mean $\pm$ SD, \%)
 \\ \hline
Baseline 
& \begin{tabular}[l]{@{}l@{}} 
Alpha F7, Theta P4, Theta Pz, Theta Oz\\ 
Theta T8, Alpha Cz, Beta F7, Alpha T7
\end{tabular} 
& 50 $\pm$ 6 
& 58 $\pm$ 8 
\\ \hline
Connectivity
& \begin{tabular}[l]{@{}l@{}}
Theta Oz, Beta P4 (PLV), Theta P3 (PLV)\\ 
Beta Pz (PLV), Alpha F7, Alpha Pz (PLV)\\
Theta P4, Alpha P4 (PLV)
\end{tabular}
& 78 $\pm$ 8
& 78 $\pm$ 8
\\ \hline \hline  
\multicolumn{4}{l}{
\begin{minipage}[t]{\textwidth}
\raggedright
\textbf{Note:} Ranked Features = Spectral Power Features \& Connectivity (PLV) Features. Ranked Features are listed in descending order of importance. PLV = Phase Locking Value (Connectivity), SD = Standard Deviation.
\end{minipage}
}
\end{tabular}
}
\label{table:1}
\end{table*}

\subsection{Feature Extraction}
\subsubsection{Spectral Power}
EEG spectral power calculation involves transforming EEG data from the time domain to the frequency domain using the Fast Fourier Transform (FFT) \cite{makeig2001}. This allows us to analyze the distribution of signal power across different frequency bands. In this study, spectral power was calculated over the entire 300-second task duration for each trial. Since all trials were of uniform length, this method facilitated a direct comparison of EEG data across trials. Mathematically, the spectral power in the \textit{i}-th frequency band was computed as: 

\begin{equation}
P_i = \log \sum_{\omega \in \text{band } i} \text{PSDs}(\omega), \quad i=\theta,\alpha,\beta.
\end{equation}

where, \(P_i\) represents the log-transformed spectral power for a specific frequency band \(i\) and \(\text{PSDs}(\omega)\) refers to the Power Spectral Density (PSD) values at each frequency \(\omega\) within the specified frequency band \(i\). 

This resulted in a total of 42 spectral features (14 channels $\times$ 3 frequency bands). To account for individual differences, the spectral power values were normalized. This was achieved by calculating the mean spectral power for each frequency band during the baseline period and subtracting these baseline values from the spectral power measurements obtained during the task. 

\subsubsection{Phase Locking Value (PLV)}
PLV is a metric of functional connectivity between different brain areas, which reflects coordinated and synchronized neural activity \cite{guan2022, thatcher2012}. Similar to the spectral power analysis, we calculated PLV values for each fronto-parietal channel pair over the entire 300-second task duration for each trial. The analysis was performed using the MNE-Python package, specifically the $spectral\_connectivity\_time$ function. This function computes the PLV between channels $k$ and $l$ over a sequence of time instances $t = \{t_1, t_2, ..., t_n\}$ as: 

\begin{equation}
PLV_{k,l} = \langle e^{j(\phi_k(t) - \phi_l(t))}\rangle
\end{equation}

\noindent in which $\langle \cdot \rangle$ is the mean over the time sequence, and $\phi_k$ and $\phi_l$ are the phases of channels $k$ and $l$ \cite{lachaux1999,pei2021}. 

PLV values range from zero (indicating random or uncorrelated phase) to 1 (representing perfect phase synchronization). It is important to note that PLV is an undirected measure, ensuring symmetry \textit{(PLV(s1, s2) = PLV(s2, s1))} \cite{lachaux1999}. 

The Morlet wavelet, a continuous wavelet transform method, was applied in conjunction with techniques for aggregating connectivity scores across different frequency bands and epochs \cite{gramfort2013}. The aggregation of connectivity scores across frequency bands was achieved using the $average\_freqs$ parameter. The output includes the median frequencies of each band, contributing to a comprehensive assessment of connectivity patterns. The $average\_epochs$ parameter facilitates the aggregation of connectivity scores over the entire trial. 

The PLV values were computed for each combination of fronto-parietal pairs. With 5 channels in either region, this yielded 5 PLV values per channel in the frontal and parietal regions. Averaging was employed to provide a consolidated representation of each channel's connectivity with the opposing frontal or parietal brain area. This resulted in a total of 10 PLV features, i.e., one (averaged) PLV feature per channel (see Figure \ref{fig:bciloop}). This computation was completed for each of the three frequency bands of interest, i.e., theta, alpha, and beta bands. As a result, a total of 30 PLV features were obtained in aggregate. 

Following the approach outlined by \cite{handy2009}, task-irrelevant synchronization was removed by normalizing the PLV values. Specifically, this process involved calculating the mean PLV value in each electrode pair and frequency band during the baseline period and then subtracting this mean from the corresponding PLV measurements in the test phase. 

\subsection{Feature Selection}
In order to enhance the model's performance and mitigate the curse of dimensionality, we employed the $sklearn.feature\_selection.RFE$ method along with a $LinearSVC$ estimator, in line with best practices demonstrated in prior studies \cite{taheri2023, kakkos2019}. This involved conducting RFE with 30,000 iterations to thoroughly explore the feature space. After earlier optimization stages, it was determined that selecting 8 features yielded the best performance for this specific problem, which were subsequently used for training. 

After the feature selection process, only a specific subset of features from each feature category was utilized during model training. Specifically, this study compared two models: a Baseline model that employed a subset of relative spectral power features (selected from 42 features), and an enhanced Connectivity model that integrated a selected subset of spectral power features with additional fronto-parietal connectivity features (selected from 42 spectral features + 30 PLV features). 

\subsection{Classification Models}
Previous attempts to create workload classifiers found success using linear discriminant analysis, SVM, or ensemble approaches involving multiple ML models \cite{arico2017, kakkos2019, taheri2023}. 

This study implemented an ensemble approach in line with the framework proposed in \cite{taheri2023}, where a stacked classifier was trained using the subset of features selected by RFE. The stacked classifier consisted of three base models—Random Forest, Logistic Regression, and SVM—using implementations from $sklearn$, with an additional SVM serving as the meta-model. Additionally, hyperparameter optimization was performed using a grid search method from $sklearn$ to fine-tune the models for improved results. 

The grid search fine-tuned the hyperparameters for the Baseline and Connectivity models, with most parameters remaining similar across models. Key differences included the number of estimators for Random Forest (100 for the Baseline model, 50 for the Connectivity model), the Logistic Regression regularization strength (1.0 for the Baseline model, 0.01 for the Connectivity model), and the SVM regularization parameter (within the stacked classifier: 0.01 for the Baseline model, 10 for the Connectivity model). 

Both models used unrestricted tree depth, a minimum of 2 samples to split nodes, and 1 sample per leaf for Random Forest, with bootstrapping enabled. Logistic Regression applied liblinear as the solver. The SVMs used a linear kernel with scale gamma, and the meta-model SVM was optimized with a regularization parameter of 10.
These hyperparameter choices were instrumental in achieving optimal performance for the respective models. While minor adjustments may cause slight performance variations, the impact is likely minimal and dataset-dependent. 

\subsection{Model Evaluation}
A 80-20 train-test split with 8-fold cross-validation (using $sklearn.model\_selection.StratifiedKFold$) was employed on the combined (VR and Desktop) dataset for model evaluation. The performance of the model was assessed using mean accuracy and mean F1-score computed over the folds. Additionally, standard deviations were calculated to gauge the variance in model performance across the folds. 

\section{Results}
The performance metrics and the selected features for the Baseline and Connectivity models are summarized in Table \ref{table:1}. 

The Baseline model achieved a mean accuracy of 50\% (±6\%) and a F1-score of 58\% (±8\%), while the Connectivity model showed a notable improvement, with a mean accuracy of 78\% (±8\%) and a F1-score of 78\% (±8\%). 

The RFE algorithm selected a subset of spectral power features that overlapped between the Baseline model and the Connectivity model. These overlapping spectral power features encompassed channels Oz and P4 in the theta band, and channel F7 in the alpha band. The selected PLV features in the Connectivity model were 1) the (averaged) connectivity between channel P4 and the frontal cortex in the alpha and beta bands, 2) the (averaged) connectivity between channel Pz and the frontal cortex in the alpha and beta bands, and 3) the (averaged) connectivity between channel P3 and the frontal cortex in the theta band. Additionally, the following spectral power features were identified by the RFE algorithm in the Baseline model: channel Pz and T8 in the theta band, channel Cz and T7 in the alpha band, and channel F7 in the beta band. 

\section{Discussion}
The current study aimed to add to the existing BCI literature on developing effective workload classifiers that can be used in virtual flight training environments, including both VR and Desktop-based simulations. Specifically, we examined the predictive power of EEG fronto-parietal connectivity in workload prediction by comparing two models: a Baseline model utilizing only spectral features and a Connectivity model incorporating both spectral and connectivity (i.e., PLV) features. Our results showed that the incorporation of fronto-parietal connectivity features led to a significant improvement in model performance, indicating their additive value in workload estimation, at least in our sample of novice participants who conducted simulated flight tasks. 

Additionally, the feature selection process sheds light on brain areas and EEG features that contributed the most to the workload classification. 
Connectivity (PLV) features in the alpha and beta frequency bands emerged as the most significant, aligning with findings in \cite{kakkos2019}, which also emphasize these bands. These findings support the increasing recognition of connectivity features across different frequencies as valuable predictors for assessing workload in diverse flight training tasks. 

It is important to note that the Connectivity model selected PLV features involving (averaged) channel P3 (located at the left parietal region) with frontal channels, (averaged) channel Pz (located at the centro-parietal region) with frontal channels, and (averaged) channel P4 (located at the right parietal region) with frontal channels. This could reflect the potential dominance of parietal cognitive processing over fronto-parietal connectivity in cognitive workload during simulated flight. 

\subsection{Limitations}
The current study presents a number of limitations that should be considered when interpreting the results. These include the small sample size, class imbalance, combining EEG data across two conditions, the lack of flight experience among participants, and the subjective nature of NASA-TLX ratings which were used as classification labels. In the following, for each of the above limitations, we discuss how the study applied a mitigation strategy to maintain the robustness of the results. 

First, the relatively small sample size and the class imblanace in the data could have caused fluctuations in classifiers' performance, particularly skewing the model outcome for the minority class \cite{roy2019}. To deal with this issue, baseline correction was applied to control for individual EEG signal differences. Additionally, we employed stratified K-fold cross-validation to ensure each fold maintained the same proportion of observations from a given class. A median split of NASA-TLX scores was also applied to balance the classes, although this approach is not without its limitations. 

To further increase the small sample size, we combined EEG data from the two experimental conditions (VR and Desktop). However this introduced another limitation in the study. Disparities in simulation fidelity between the two conditions—specifically differences in the fields of view—could have impacted the quality of EEG signals, potentially influencing subsequent information processing \cite{hagedorn2023}. Moreover, since each participant completed both conditions, this raised concerns about ``carryover effects" (where completing two similar tasks consecutively could influence the participant's performance \cite{sutter2011}), and the risk of ``identity confounding" (where the algorithm might learn biased patterns based on individual characteristics due to the use of two or more data points of the same person). To address this limitation, we conducted a post-hoc analysis to assess the above-mentioned concerns (reported in Appendix~\ref{AppendixA}). The analysis revealed no significant evidence of "carryover effects" nor ``identity confounding", supporting the validity of our approach in merging the datasets. 

Another limitation of this study was the lack of flight experience among participants, which can limit the applicability of the findings to adaptive systems for trained pilots and the generalization of the results to real-world applications. However, previous research, including a recent study conducted by our team, demonstrated that workload classification can achieve 80\% accuracy with military pilots in using EEG spectral power features during VR flight tasks \cite{vanweelden2024a}. By incorporating connectivity features, the pipeline introduced in this study achieves approximately 30\% higher classification accuracy, emphasizing its potential for real-world applications with trained pilots. 

The final limitation of the current study was the usage of NASA-TLX scores as workload labels and the application of a median split to balance the classes. Although over many years, the NASA-TLX has consistently proven to be a robust measurement of cognitive workload, especially in aviation research \cite{hart2006}, multiple studies have identified potential limitations stemming from its subject-specific variability \cite{chaibubneto2019, grier2015}. For instance, \cite{bustamante2008} found that the NASA-TLX lacked scalar invariance, which may lead to biased comparisons between individuals or groups. Furthermore, \cite{hart2006} and \cite{grier2015} emphasize the importance of controlling for context effects, as the different conditions (VR vs. Desktop) and flight tasks (medium turn vs. speed change) could have influenced the NASA-TLX scores. To address the potential bias introduced by the median split, which assumes a universal threshold for NASA-TLX scores, we replaced the raw NASA-TLX scores with residuals from a mixed-effects regression model. This model controlled for individual variability, context effects (e.g., condition type and flight task), and baseline differences, preserving data integrity and enhancing the robustness of the analysis (reported in Appendix~\ref{AppendixB}). 

\subsection{Future Work}
In future work, exploring alternative models like non-linear prediction models may enhance workload prediction \cite{Nittala2018, antoine2022}. More complex feature subsets and configurations of PLV values, including temporally defined features, could provide finer-grained insights. Comparing different connectivity metrics such as Parietal Directed Coherence (PDC) \cite{sun2014, baccala2001}, and PLI \cite{kakkos2019, pei2021} in relation to PLV may offer a further understanding of neural dynamics representing different workload levels. 

Whilst this study did not employ data augmentation techniques to address class imbalance due to concerns about increased risk of overfitting \cite{roy2019}, future work could explore oversampling methods, such as the Synthetic Minority Over-sampling Technique (SMOTE) \cite{chawla2002}, Safe-Level SMOTE \cite{gosain2017}, or weighted loss functions \cite{fernando2022}, which may more effectively manage class imbalance while maintaining the data’s original structure. 

Building on the current study, future work can enhance workload classification models by integrating objective performance metrics, such as behavioral measures, task accuracy, or completion time, alongside subjective labels like NASA-TLX. This approach will reduce the over-reliance on subjective labels \cite{chaibubneto2019, grier2015} and offer a more comprehensive understanding of workload, enabling triangulation of workload classification and ensuring that both self-reported experiences and objective performance are effectively captured. 

Another important direction for future research and innovation involves greater use of VR simulations for EEG data collection in aviation studies. Due to the challenges of capturing EEG data in real-world flight scenarios, high-fidelity VR environments offer an immersive, controlled setting for assessing pilot workload \cite{vanweelden2021}. Besides, integrating workload classifiers within immersive, high-fidelity VR environments could greatly enhance pilot training by providing real-time feedback based on cognitive workload. Such tools would allow for dynamic adjustments to training scenarios, catering to the individual cognitive states of pilots. 

Finally, it is recommended to evaluate whether our findings can be replicated with novices or further explored using experienced pilots. This could extend beyond VR flight simulations to include other training environments, such as cockpit simulators or real flight. By pursuing these avenues in future research, a more comprehensive understanding of workload during flight training and its underlying neural correlates can be achieved, ultimately contributing to the development of more accurate and robust real-time neuro-adaptive systems. 

\section{Conclusion}
The aim of this study was to evaluate the potential enhancement of workload-monitoring BCIs in virtual flight simulations through the incorporation of EEG connectivity features. A notable improvement in the workload classification accuracy was observed compared to a baseline model. Moreover, the feature selection process provided valuable insights into the prioritized features, notably emphasizing the importance of the phase synchronization between frontal and parietal regions. This research provides a first step in introducing PLV as a feature for EEG-based workload classification in VR aviation training. Future investigation using larger and more homogeneous EEG datasets from pilots is crucial to clarify the true contribution of PLV features in the evaluation of mental workload and performance in simulated flight. 

\IEEEtriggeratref{47}
\bibliographystyle{IEEEtran}
\bibliography{biblio}

\def\BibTeX{{\rm B\kern-.05em{\sc i\kern-.025em b}\kern-.08em
    T\kern-.1667em\lower.7ex\hbox{E}\kern-.125emX}}

\appendix 

In this appendix, we present a brief post-hoc study aimed at further validating our primary findings. This study is structured around three specific goals. The three primary goals of this post-hoc analysis are: (1) to evaluate potential confounding effects related to the carryover effect and ``identity confounding"; (2) to reduce between-subject variability in NASA-TLX scores by incorporating residuals from a mixed-effects model; and (3) to justify the use of a subset of NASA-TLX subscale ratings in workload analysis.

\subsection{Evaluation of Carryover Effect and ``Identity Confounding"}
\label{AppendixA}
To validate the merging of datasets from the Desktop and VR conditions, an evaluation was conducted to determine whether significant differences existed between these conditions. Given that each participant experienced both conditions in a sequential manner, potential carryover effects \cite{sutter2011} were also examined. A paired t-test was performed to assess any ``identity confounding" between both conditions and to account for potential order effects due to the sequential presentation of conditions. ``Identity confounding" refers to a situation where the algorithm might learn biased patterns based on individual characteristics due to the use of two or more data points from the same person \cite{chaibubneto2019}. The test was considered valid, as the data adhered to the normality assumption, as confirmed by the Kolmogorov-Smirnov (KS) test. 

The KS test indicated no significant difference, with \textit{D} = .13 and \textit{p} = .36. No significant difference was found between the workload values of Desktop (\textit{M} = 47.00, \textit{SD} = 17.87) and VR (\textit{M} = 50.85, \textit{SD} = 20.59), \textit{t}(48) = -1.81, \textit{p} = .07. Also the KS test indicated no significant difference between exposures, with \textit{D} = .11 and \textit{p} = .55. No carryover effect was observed, meaning that the order in which conditions were presented to each subject did not result in a significant difference in workload levels between the first exposure (\textit{M} = 47.74, \textit{SD} = 19.61) and second exposure (\textit{M} = 50.11, \textit{SD} = 19.07), \textit{t}(48) = -1.09, \textit{p} = .28. These findings suggest that neither order effects nor condition-specific workload differences confounded the data, supporting the decision to merge the VR and Desktop conditions for the classification. 

It is worth noting that a similar paired \textit{t}-test was conducted in previous research using this dataset \cite{vanweelden2022comparing}. 
However, the values of this analysis differ slightly due to variations in sample sizes, as some participants were excluded as outliers in this study. 

\subsection{Reducing NASA-TLX Between-Subject Variability by Incorporating Residuals from a Mixed-Effects Model}
\label{AppendixB}
NASA-TLX was originally designed for within-subject comparisons, allowing researchers to assess workload variations across different conditions rather than make direct between-subject comparisons \cite{grier2015}. This design introduces challenges when comparing workload scores across participants due to individual differences in perception, rating tendencies, and baseline biases. As a result, NASA-TLX scores may reflect personal biases or experiences rather than objective workload differences, complicating their use in between-subject analyses. This variability can obscure meaningful comparisons and reduce classification accuracy in workload studies, as subjective factors dominate over objective workload differences. A meta-analysis \cite{grier2015} underscores these challenges, noting variability across subjects and domains. The median split used to balance classes in this study does not fully address these biases and risks misclassification when universal thresholds are applied to subjective data. 

To address these challenges, a mixed-effects model was applied to control for between-subject variability. This analysis was conducted using the \textit{statsmodels} package, specifically utilizing the \textit{mixedlm} function from the \textit{statsmodels.formula.api} module \cite{seabold2010statsmodels}. The Four-Scale workload score was modeled as the dependent variable, with Condition Type and Flight Task as fixed effects to control for potential confounds. Participants were included as random effects to account for individual differences in baseline workload. The model featured a random effect implemented by a varying intercept for each participant, capturing deviations from the overall mean workload. 

The analysis revealed no significant effects for Condition Type (\textit{Coef.} = 3.87, \textit{SE} = 2.16, \textit{z} = 1.79, \textit{p} = .07) or Flight Task (\textit{Coef.} = .14, \textit{SE} = 2.16, \textit{z} = .07, \textit{p} = .94). However, substantial individual variability was observed, with a large group variance (\textit{Group Variance} = 258.82, \textit{SE} = 8.58). These findings highlight the notable impact of participant-specific factors on workload ratings, underscoring the need to adjust for these differences. 

To mitigate these effects, residuals from the mixed-effects model were used as replacement workload labels, a strategy similar to that in \cite{moqaddasi2019}, which has shown improved results compared to other regression methods. Unlike normalization or scaling methods that apply uniform adjustments across all participants, residuals from a mixed-effects model account for both fixed effects and random effects (such as individual participant differences). By removing the subject-specific variance, residuals preserve meaningful variations linked to conditions and tasks while adjusting for individual baseline differences. This method avoids the oversimplification of other approaches and provides a more robust, interpretable way to compare workload across subjects. 

Residuals represent deviations from predicted values based on fixed effects, effectively removing participant-specific baseline differences. This adjustment enables a refined comparison of workload across subjects, reducing the influence of individual biases. 

Consequently, EEG segments were categorized into two distinct classes based on the residuals from the mixed-effects model and the median split: low workload, corresponding to residual values at or below .612, and high workload, corresponding to residual values above .612. 

\subsection{NASA-TLX Subscale Subsets in Workload Analysis}
\label{AppendixC}
The use of NASA-TLX subscale ratings or partial subscale aggregation is also a common practice, as they can provide valuable insights beyond a single overall workload score \cite{hart2006, hertzum2021, tubbs-cooley2018, braarud2020}. Research has demonstrated that task-specific studies and conditions often reveal significant correlations between the subscales \cite{hart2006}, suggesting that they collectively measure different facets of the same underlying construct. For instance, analyzing subscale ratings allows designers to pinpoint specific sources of workload or performance issues more precisely. Moreover, empirical evidence from various research fields supports the use of different sets of subscales to enhance workload prediction accuracy \cite{hertzum2021, tubbs-cooley2018, braarud2020}. 

To enhance the granularity and relevance of workload assessment, the study evaluated all possible combinations of NASA-TLX subscale subsets to determine the most effective combination for predicting workload. The analysis identified that the Four-Scale subset including Mental Demand, Physical Demand, Performance, and Effort provided the highest predictive classification accuracy, making it the preferred set for analysis. In computing the average score for this subset, equal weights were applied to each subscale \cite{hertzum2021}. 

\subsection{Summary}
The merging of two datasets from VR and Desktop conditions was not confounded by systematic order effects or condition-specific workload differences. 

Additionally, an important observation is the significant role of participant-specific baseline differences on workload scores. To account for these individual differences, an approach was employed where the workload labels were replaced by residuals derived from the mixed-effects model.

Finally, the analysis revealed that the Four-Scale NASA-TLX subset, which includes Mental Demand, Physical Demand, Performance, and Effort, provided the highest predictive accuracy in distinguishing workload levels. 

\clearpage



\end{document}